\documentclass[aps,prb,twocolumn,superscriptaddressgroupaddress,floatfix]{revtex4-2}
\usepackage{amssymb}
\usepackage{graphicx}
\usepackage{times}
\usepackage{amsmath}
\usepackage{amsthm}
\usepackage{amsfonts}
\usepackage[T1]{fontenc}
\usepackage[latin9]{inputenc}
\usepackage{array}
\usepackage{multirow}
\usepackage{color}
\usepackage{bm}
\usepackage{color}

\usepackage{bbm}
\usepackage{hyperref}
\usepackage{babel}
\usepackage[mathscr]{euscript}
\usepackage{float}
\usepackage{hyperref}
\hypersetup
{	colorlinks,%
	citecolor=green,%
	linkcolor=red,%
	urlcolor=blue%
}
\allowdisplaybreaks[4]

\begin{document}
\title{Energetic comparison of exciton gas versus electron-hole plasma in a bilayer two-dimensional electron-hole system}
\author{Yi-Ting Tu}
\author{Seth M.\ Davis}
\author{Sankar Das Sarma}
\affiliation{Condensed Matter Theory Center and Joint Quantum Institute, Department of Physics, University of Maryland, College Park, Maryland 20742, USA}

\begin{abstract}
  We study the zero-temperature phase diagram of a symmetric electron-hole bilayer system by comparing the ground state energies of two distinct limiting cases, characterized by an electron-hole plasma or an exciton gas, respectively.
  For the electron-hole plasma, the random phase approximation is used; for the exciton gas, we consider three different approximations: the unscreened Coulomb interaction, the statically screened one, and the dynamically screened one under the plasmon-pole approximation.
  Our results suggest that the exciton gas is stable at small layer separation. However, static screening in general suppresses the formation of excitons, and dynamic screening gives different results depending on the representative energy scale we used in the plasmon-pole approximation.
  We conclude that energetic considerations alone are very sensitive to the approximation schemes, and the phase diagram of the system may depend crucially on exactly how the electron-hole attraction is treated in the theory.
  For very small and very large densities, however, all our approximations show the exciton gas to have lower energy than the plasma.

\end{abstract}

\maketitle

\section{Introduction}

The idea that particles and holes in a solid state system could pair up to form \textit{excitons}, hydrogen-atom-like bound states, and consequently give rise to novel states of matter, has a long and complicated history in the condensed matter literature, spanning almost a century~\cite{Frenkel1931,Lozovik1975,Lozovik1976,DasSarma2000,Eisenstein2004}.
However, due to the fast rate of recombination of the constituent particles, stable exciton-based phases have never been realized in traditional, three-dimensional solid state systems.
On the other hand, the advent of two-dimensional heterostructures, in which exciton are stabilized due to the electron and hole being confined to different layers, thus substantially reducing recombination~\cite{Lozovik1975,Lozovik1976,Eisenstein2004}, has caused a renaissance in the investigation into excitons.
Consequently, 2D exciton systems are now being produced in the lab \cite{Wang2019,Ma2021,Davis2023} and their properties are being explored.
Exciton phases could be novel, useful states of matter in solid state physics, possibly with some eventual applications such as in optics and optoelectronics~\cite{Mueller2018,Anantharaman2021}.

One central question is the phase diagram of such a bilayer system, where one layer is doped with electrons and the other with holes. There are many possible phases, including an electron-hole (eh) plasma, an exciton gas/liquid, an exciton condensate, or a Wigner crystal phase~\cite{Joglekar2006}.
The phase diagram has been studied extensively using several variations of the Monte Carlo method~\cite{DePalo2002,Filinov2003,Schleede2012,Maezono2013,Sharma2016,Sharma2018,DePalo2023}, and also using exact diagonalization~\cite{Kaneko2013}, density-matrix renormalization group~\cite{Vu2023}, diagrammic calculation~\cite{Babichenko2018}, and mean-field theory~\cite{Zhu1995,Wu2015,Trushin2022}.
However, the results do not all agree across the various approaches, and different methods and assumptions yield a huge range of possible transition temperatures $T_c$ into the condensate phase.
In addition, numerical approaches require some uncontrolled assumptions, making the comparison among various approximations difficult.
For example, in the variational Monte Carlo approaches, the variational wave function already assumes the formation of excitons; in the mean-field approach, the Coulomb interaction is often taken to be the unscreened one, not even the screened interaction in the random phase approximation (RPA) (but the latter has been used in a slightly different context for a one dimensional electron-hole system~\cite{DasSarma2000}).
A key unresolved question is whether the bilayer system allows a stable electron-hole plasma ground state for some values of the system parameters (i.e.\ layer separation and layer density) or the excitonic system is always the ground state.
It is especially interesting to see if an exciton condensate phase can exist at a high temperature, which is a key open theoretical question (not considered in the current work, where we compare only the plasma and exciton gas phases).

In the graphene literature, the problem of exciton condensation has been studied using different models of the Coulomb interaction~\cite{Min2008,Kharitonov2008,Lozovik2008,Abergel2012,Sodemann2012,Abergel2013}, where it is found that the result depends crucially on the model of screening.
In particular, while the unscreened model gives a reasonable $T_c$ for the excitonic condensate, the statically screened model suppresses it exponentially, ruling out bilayer exciton condensation at any experimentally relevant temperature.

In this work, we address the question of the stability of the exciton gas by considering the 2D bilayer electron-hole system in two extreme limits: one in which the system is an electron-hole bilayer plasma and another in which it is a bilayer exciton gas. Directly comparing the ground state energy per particle of these two distinct systems (but with the same parameters) gives a hint as to which one is energetically favored in which regime, with the caveat that we are carrying out an energetic comparison between two possible,  but qualitatively-distinct regimes.
In the first limit, we assume the system is an interacting electron-hole plasma, with electrons and holes in separate 2D layers. We use the RPA to calculate the ground state energy, which is exact in the high-density limit. In the competing case, we assume that non-perturbative effects of the particle-hole attraction have lead to the creation of interlayer excitons. In this limit, we may simply take the ground state energy per particle to be the ground state energy of a single exciton, which we get directly from the Schr\"odinger equation. By reconsidering the latter calculation with a screened interaction, we may study how screening affects the stability of the exciton gas. In this paper, we do this comparison for a symmetric 2D bilayer system with parabolic band dispersions, and plot the resulting phase diagram with respect to the density parameter $r_s$ and the layer separation $d$. We do this for different models of screening---both static and dynamic screening cases are considered. For the latter, we use the plasmon-pole approximation with the external frequency fixed to a representative energy scale, which should be a reasonable approximation for dynamical screening. Essentially, we solve the excitonic Bethe-Salpeter equation using the unscreened or the screened interlayer electron-hole Coulomb interaction.

By carrying out an energy comparison between these extreme cases (bilayer electron-hole plasma and bilayer exciton gas), we construct a qualitative phase diagram for which of these two phases is energetically lower as a function of 2D electron/hole density and layer separation.
We find that, while excitons are preferred at small $d$ in the unscreened model, static screening suppresses the formation at all $d$ (except for very small or very large $r_s$).
Dynamical screening, on the other hand, gives a result either similar to the unscreened case or the statically screened case, depending on what representative energy scale we use.
Interestingly, for small $r_s$ (i.e.\ high density), the exciton gas always has lower energy compared with the electron-hole plasma for all layer separations.  The same seems to be true generically for very large $r_s$ (i.e.\ very low density).

\section{Theory}

We consider a two-dimensional, two-layer system, with one layer containing (spinless) electrons and the other containing (spinless) holes with no interlayer tunneling. (Inclusion of spin is an unnecessary complication which would not change our conclusion since the Coulomb interaction is spin independent.) The layers are separated by a distance $d$, and are coupled by the electron-hole Coulomb interaction. We take both the electrons and holes to have parabolic dispersion with mass $m$, and all particles interact with each other via the Coulomb interaction. We consider the balanced situation with equal electron-hole density in the layers. To approximate effects of charge-gating, each layer has a background charge that neutralizes its total charge. (That is, each layer is a 2D ``jellium'' system---in experimental systems, there are external gates controlling the layer densities and these gates act as the neutralizing background.) Both layers have $N$ particles, and are described by the usual dimensionless density parameter $r_s$, which gives a rough estimate of the ratio of potential and kinetic energy in the system. Explicitly, the density parameter (which is also the coupling constant for Coulomb interaction) $r_s$ is defined by
\begin{align}
    r_s = \frac{S}{\pi a_B^2 N},
\end{align}
where $S$ is the 2D area of the system.
Above and in the rest of the manuscript, we use Gaussian units with $\hbar=1$.
Also, the Rydberg $\mathrm{Ry}=me^4/2$ can be used as a natural energy unit and the Bohr radius $a_B=1/me^2$ the natural length unit.

The Hamiltonian is
\begin{multline}
  H = \sum_{\mathbf{k},\sigma} \frac{k^2}{2m} c^\dagger_{\mathbf{k},\sigma} c_{\mathbf{k},\sigma} \\
  + \frac{1}{2}\frac{1}{S} \sum_{\mathbf{q}} \sum_{\mathbf{k},\sigma} \sum_{\mathbf{k}',\sigma'}
  V_{\sigma,\sigma'}(q) c^\dagger_{\mathbf{k}+\mathbf{q},\sigma} c^\dagger_{\mathbf{k}'-\mathbf{q},\sigma'} c_{\mathbf{k}',\sigma'} c_{\mathbf{k},\sigma} \\
  + \frac{N}{S} \sum_{\mathbf{k},\sigma} \sum_{\sigma'} (-V_{\sigma,\sigma'}(q=0)) c^\dagger_{\mathbf{k},\sigma} c_{\mathbf{k},\sigma} \\
  + \frac{1}{2} \frac{{N}^2}{S} \sum_{\sigma} \sum_{\sigma'} V_{\sigma,\sigma'}(q=0).
\end{multline}
The four terms are, respectively: kinetic energy of the particles, interaction between two particles,  interaction between a particle and a charge-compensating background, interaction within the charge-compensating backgrounds.
Here, $\sigma$ is the layer index indicated by $+$ (holes) and $-$ (electrons). The Coulomb interaction is
\begin{equation}
  V_{\sigma,\sigma'} =
  \begin{cases}
    +\frac{2\pi e^2}{q} & \sigma=\sigma', \\
    -\frac{2\pi e^2}{q}e^{-qd} & \sigma\neq\sigma'.
  \end{cases}
\end{equation}
All interlayer tunneling effects are neglected, assuming the two layers to be coupled only by Coulomb interaction and not by single particle tunneling.

Note that if we separate the $q=0$ and $q\neq 0$ part of the interaction between electrons and holes, then all the $q=0$ terms will cancel exactly the terms involving the charge-compensating background.
Therefore, we are left with an equivalent Hamiltonian with only the kinetic and the electron-hole interaction terms but with $V_{\sigma,\sigma'}(q=0)$ redefined to be zero,
\begin{multline}
  H = \sum_{\mathbf{k},\sigma} \frac{k^2}{2m} c^\dagger_{\mathbf{k},\sigma} c_{\mathbf{k},\sigma} \\
  + \frac{1}{2}\frac{1}{S} \sum_{\mathbf{q\neq 0}} \sum_{\mathbf{k},\sigma} \sum_{\mathbf{k}',\sigma'}
  V_{\sigma,\sigma'}(q) c^\dagger_{\mathbf{k}+\mathbf{q},\sigma} c^\dagger_{\mathbf{k}'-\mathbf{q},\sigma'} c_{\mathbf{k}',\sigma'} c_{\mathbf{k},\sigma}. \\
\end{multline}

Since the zero-momentum component of the effective interaction is zero, all tadpole diagrams including a zero-momentum interaction line will vanish.

We will also make use of the ``excitonic units''~\cite{Tan2005}. These are defined as
\begin{align}
    a_B^*=2a_B,
    \\
    1\;\mathrm{Ry}^*=\frac{1}{2}\;\mathrm{Ry}.
\end{align}
We will use $a_B^*$ as the unit of length and $\mathrm{Ry}^*$ as the unit of energy throughout this paper (unless otherwise noted).

\subsection{Electron-hole plasma}
We calculate the ground state energy per electron/hole assuming that the system is an electron-hole plasma, in excitonic units,
\begin{equation}
  \frac{E}{N} = \frac{8}{r_s^2} - \frac{64}{3\pi r_s} + \frac{E_2^\text{(b)}}{N} + \frac{E_\text{corr}}{N}
\end{equation}
\begin{figure}
  \includegraphics[trim=0 0 0 0, clip, scale=0.7]{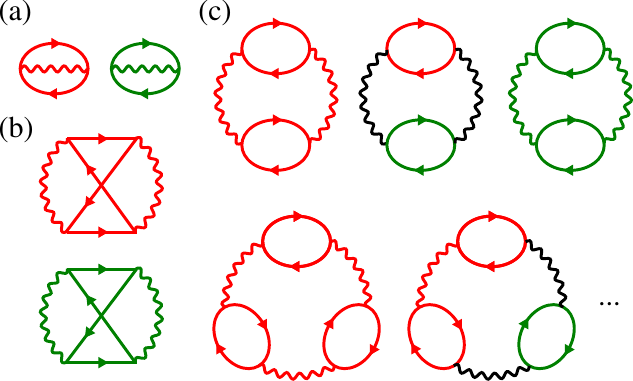}
  \caption{The Feynman diagrams used in the calculation of the eh plasma energy. Red solid line: electron; green solid line: hole; red wavy line: Coulomb interaction between holes; green wavy line: Coulomb interaction between electrons; black wavy line: interlayer Coulomb interaction. (a) The Fock energy. (b) The second order exchange correlation energy $E_2^\text{(b)}$. (c) The RPA correlation energy $E_\text{corr}$, which includes all ring diagrams of second or higher orders.}
  \label{fig:diagrams}
\end{figure}
where the four terms are the kinetic energy, the Fock (i.e.\ the exchange) term, the second-order exchange energy, and the ring contribution (RPA) of the correlation energy, respectively~\cite{Rajagopal1977} (see the diagrams in Fig.~\ref{fig:diagrams}).
As mentioned above, the Hartree terms in the expansion vanish due to cancellation with the background charge.
Note also the excitonic electron-hole attraction inside the polarization bubbles is not explicitly included in the plasma energetics since these are vertex corrections, and therefore, higher order effects beyond RPA, where only the bare ring/bubble diagrams are retained.

We include the second-order exchange energy because the divergence in the RPA series starts at third-order in $r_s$. This energy is given by~\cite{Rajagopal1977} 
\begin{multline}
  \frac{E_2^\text{(b)}}{N} = \frac{1}{\pi^3}\int d^2\mathbf{q} \int d^2\mathbf{k}_1 \int d^2\mathbf{k}_2\\
  f(k_1)f(k_2)(1-f(|\mathbf{k}_1+\mathbf{q}|))(1-f(|\mathbf{k}_2+\mathbf{q}|))\\
  \frac{1}{q|\mathbf{q}+\mathbf{k}_1+\mathbf{k}_2|}
  \frac{1}{q^2+\mathbf{q}\cdot(\mathbf{k}_1+\mathbf{k}_2)},
\end{multline}
where $f(k)=1$ if $k<1$ and zero otherwise, and is numerically evaluated to approximately $0.456\;\mathrm{Ry}$.

To calculate $E_\text{corr}$, we first write the full RPA energy (sum over all elementary rings or polarization) as
\begin{equation}
  E_\text{RPA} = F_+ + F_-
\end{equation}
\begin{equation}
  \frac{F_\pm}{S} = \frac{1}{2}\int_0^\infty \frac{dq}{2\pi}q\int_{-\infty}^\infty \frac{d\omega}{2\pi}
  \log\left[1-\left(1\pm e^{-qd}\right)\frac{2\pi e^2}{q}\Pi(\omega,q)\right]
\end{equation}
where we have defined the polarization operator~\cite{Stern1967}
\begin{equation}
  \Pi(\omega,q)=\frac{m}{2\pi}\left\{\mathrm{Re}\left[\sqrt{\left(1-i\frac{2m}{q^2}\omega\right)^2-\frac{4p_F^2}{q^2}}\right]-1\right\}.
\end{equation}
Then the (ring contribution of) correlation energy $E_\text{corr}$ is obtained from $E_\text{RPA}$ by subtracting the term in the integrand which is linear in $r_s$ (from the expansion of log).
The final expression expressed using the excitonic units and the dimensionless integration variables $\tilde q=q/p_F$, $\tilde\omega=2m\omega/(p_F q)$ (chosen for numerical stability) is

\begin{equation}
  E_\text{corr} = F_\text{corr}^+ + F_\text{corr}^-
\end{equation}
\begin{multline}
  \frac{F_\text{corr}^\pm}{N} = \frac{8}{\pi r_s^2}\int_0^\infty d\tilde q \tilde q^2\int_0^\infty d\tilde\omega\\
  g\left[-\left(1\pm e^{-\frac{4\tilde q}{r_s} d 
  }\right)\frac{r_s}{\tilde q^2}\;h\left(\frac{\tilde q}{2}+i\frac{\tilde\omega}{2}\right)\right],
\end{multline}
where we define the following functions,
\begin{equation}
  g(x)=\log(1+x)-x,
\end{equation}
\begin{equation}
  h(z)=\mathrm{Re}\left(\sqrt{z^2-1}-z\right).
\end{equation}
We obtain the ground state energy as a function of the parameters $d$ and $r_s$ by evaluating the above dimensionless integral numerically. The results are shown as the gray curves in Figure~\ref{fig:cuts}.
The characteristic features of the calculated plasma ground state energy are: (1) it is featureless and only weakly $d$-dependent (except for $d<r_s$) as a function of interlayer separation; (2) it is a strong nonmonotonic function of $r_s$ with a minimum around $r_s\sim2$ indicating a manifestly stable plasma phase for $r_s\sim2$.  These features play an important role in the phase diagram.

\subsection{Exciton gas}
We will now consider the opposite limit that assumes that the electrons and holes form well-defined excitons. We will call this limit the ``exciton gas'' and will estimate the ground state energy per particle from the calculated binding energy of the individual excitons.
We ignore exciton-exciton interactions and focus on the exciton gas, but we include through screening the effects of the carriers themselves on the exciton gas stability at an effective one particle mean field level.

\begin{figure}
  \includegraphics[trim=0 0 0 0, clip, scale=0.7]{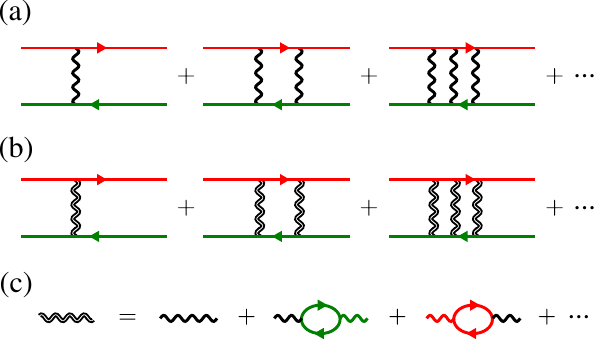}
  \caption{The diagrams for the Bethe-Salpeter equation describing the excitons. (a) Unscreened electron-hole Coulomb coupling; (b) screened Coulomb coupling; (c) the screening approximation in RPA. Here the solid lines with arrows are the electron and hole propagators in each layer (in the single exciton limit), and the wavy/double wavy lines are the unscreened and screened interlayer electron-hole coupling between the two layers with the polarization bubble being the screening function approximated in either the zero-frequency limit (``static screening'') or a fixed finite-frequency limit (``plasmon pole screening'') as described in the main text.}
  \label{fig:exciton}
\end{figure}

The exciton gas approximation (see Fig.~\ref{fig:exciton}) is strengthened by the fact that the first-order energy corrections due to residual dipole-dipole interactions are zero due to the background charge cancellation. Explicitly, the dipole-dipole interaction ($V_{DD}(q)$) between excitons is given by
\begin{align}
  V_{DD}(q) = V_{++}(q) + V_{--}(q) - 2 V_{+-}(q).
\end{align}
Since the potentials $V_{\sigma,\sigma'}$ all vanish at $q=0$ due to background charge cancellation, so does the dipole-dipole interaction.
As we are considering the zero-temperature phases, the free exciton gas is in its ground state, where all excitons have zero momentum.
Hence the first-order correction, which only considers the processes from the ground state to the ground state, vanishes.
There are higher order terms arising from all the vertex (and self-energy) corrections, but these are intractable diagrammatically within our approximation scheme, where we want to compare the plasma energy with no vertex corrections to the exciton energy with no many exciton effects.

We calculate the exciton binding energy by solving the hydrogen-like time-independent Schr\"odinger equation for the electron-hole pair (Fig.~\ref{fig:exciton}). Since we are only interested in the ground state, we assume zero angular momentum. When expressed in terms of the radial coordinate using the excitonic units, the effective exciton binding equation for our electron-hole bilayer becomes~\cite{Tan2005}

\begin{equation}
  -\frac{1}{r}\frac{\partial}{\partial r}\left(r\frac{\partial\Phi}{\partial r}\right)+2 V(r) \Phi(r) = E_X \Phi(r),
\end{equation}
where $\Phi(r)$ is the (radial) wavefunction, $V(r)$ is the interlayer Coulomb interaction, and $E_X$ is the exciton binding energy.

In the case of the unscreened Coulomb interaction, we use the radial potential
\begin{equation}
  V_\text{unscreened}(r) = -\frac{1}{\sqrt{r^2+d^2}}.
\end{equation}
In the momentum space, this is (with some normalization),
\begin{equation}
  V_\text{unscreened}(q) = -\frac{e^{-qd}}{q}.
\end{equation}

In addition to the unscreened case, we also consider the case in which the Coulomb interaction within each interlayer exciton is screened by the eh plasma background. This is equivalent to considering the question of whether or not the eh plasma will be destabilized by the formation of excitons.
One should think of this screening as an effective mean field theory motivated on physical grounds, with the goal of approximately including the effect of all the other particles in the system beyond the single exciton. Such a screening approximation can only be justified on heuristic grounds since excitons themselves do not provide effective screening as they are charge neutral.

First, we consider the statically screened Coulomb interaction, in which we approximate the potential using the long-wavelength limit~\cite{Lozovik1976, Kharitonov2008}:
\begin{equation}
  V_\text{static}(q) = -\frac{e^{-qd}}{q + 2\varkappa + \varkappa^2 (1-e^{-2qd})/q}
\end{equation}
where $\varkappa=2$ is the inverse Debye screening length.
This is basically the Thomas-Fermi or Debye screening which is density-independent in 2D.
We also consider the dynamically screened Coulomb interaction
\begin{equation}
  V_\text{dynamic}(q,\omega) = -\frac{e^{-qd}}{q \epsilon(q,\omega)}
\end{equation}
where the dielectric function $\epsilon(q,\omega)$ is obtained from the plasmon-pole approximation for the sake of simplicity
\begin{equation}
  \epsilon(q,\omega) = 1 - \frac{\omega_p^2}{\omega^2 - \omega_q^2},
\end{equation}
$\omega_p$ is the long-wavelength 2D plasmon frequency (in the unit of $\mathrm{Ry}^*$)
\begin{equation}
  \omega_p = \sqrt{\frac{4\pi N e^2 q}{m S}} = \frac{4}{r_s}\sqrt{2q},
\end{equation}
and $\omega_q$ is determined by the condition
\begin{equation}
  V_\text{dynamic}(q,0) = V_\text{static}(q).
\end{equation}
We use the approximation in which the external frequency $\omega$ is assumed to be a constant, so that we have again a static potential to solve the time-independent Schr\"odinger equation.
We try three different energy scales for $\omega$: the unscreened exciton binding energy ($\omega=-E_X$), the Fermi energy of the electron-hole plasma ($\omega=E_F$), and the plasmon frequency at the Fermi momentum of the electron-hole plasma ($\omega=E_p=\omega_p(q_F)$).
Note that similar approximations have been used in previous studies (in different contexts) and have been shown to be good~\cite{DasSarma1988,Jain1988,DasSarma1990}.
We guess that $\omega=-E_X$ might be the most relevant energy scale physically (as the scale is likely set by the process of exciton formation itself).
However, deciding which approximation is the most accurate in an experimental setup requires some more detailed microscopic theory and is beyond the scope of this work.
Therefore, we just present the results for each of the approximations in this paper.

For unscreened and statically screened cases, the ground state energy is obtained by numerically solving the Schr\"odinger equation using standard numerical integration techniques; for the dynamically screened cases, it is approximately solved using variational method with an exponential trial wavefunction proportional to $\exp(-r/a)$ with parameter $a$ (this approach avoids the highly oscillatory $V(r)$ by doing the calculation directly from $V(q)$).
Our main interest being the qualitative phase diagram, these approximations are adequate for our purpose.

\section{Results}

\begin{figure*}
  \includegraphics[trim=0 0 0 0, clip, scale=0.9]{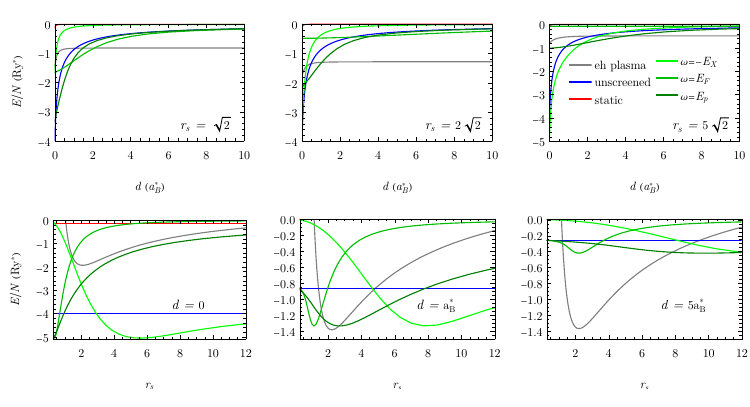}
  \caption{Ground state energy per electron/hole of bilayer electron-hole system as a function of $d$ and of $r_s$ under different assumptions. In each subfigure, the six curves represent the assumptions that the system is: electron-hole plasma (gray); exciton gas with unscreened Coulomb interaction (blue), statically screened Coulomb interaction (red), and the plasmon-pole approximated dynamically screened Coulomb interaction with external frequency $\omega$ being the unscreened exciton binding energy (light green), the Fermi energy of the eh plasma (green), and the plasmon energy at the Fermi momentum (dark green).
  For very large $r_s$, the eh plasma curves will become positive again (not shown).
  }
  \label{fig:cuts}
\end{figure*}

\begin{figure*}
  \includegraphics[trim=5 0 0 0, clip, scale=0.9]{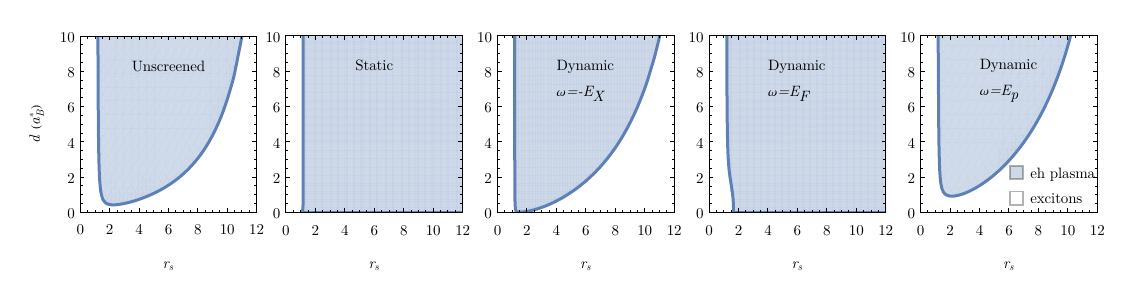}
  \caption{Phase diagrams of the bilayer electron-hole system obtained from our models. The five subfigures describe different screening models of the interlayer Coulomb interaction used in the calculation of excitonic binding energy.
  The ``phase'' here is determined by comparing the ground state energy assuming the two extremes, so the blue (white) region corresponds to where the eh plasma energy per electron/hole is lower (higher) than the exciton binding energy.
  } 
  \label{fig:phase}
\end{figure*}

Figure~\ref{fig:cuts} shows the calculated $E/N$, the ground state energy per electron/hole (equivalent to per exciton for the exciton gas) under various assumptions or various parameters, particularly as a function of density $r_s$ and interlayer separation $d$.
We can see that the statically screened free exciton model shows much smaller (less negative) binding energies than the unscreened model, as expected.
For the dynamically screened model, although it is usually between the unscreened and statically screened case, it sometimes shows a larger (more negative) binding energy than the unscreened model, manifesting the anti-screening effect which is well-known in the context of dynamical screening (because the dielectric function could become less than unity depending on the parameters).
Note that in experiments such as transition metal dichalcogenide (TMD) \cite{Wang2019,Ma2021}, $r_s$ is tunable, with the maximum being the order of $r_s\sim 10$. Therefore, the parameter range in our calculation is experimentally relevant.

For a given exciton screening assumption (unscreened, static, $\omega=-E_X$, $\omega=E_F$, or $\omega=E_p$), we can compare the corresponding $E/N$ (a colored line) with that of the eh plasma (the gray line).
If the latter is larger (more negative), it means that the eh plasma is stable in our theory; if the former is smaller, it means that the exciton phase is more stable (or at least that the eh plasma phase will be destabilized by the formation of excitons).
The resulting phase diagram is shown in Figure~\ref{fig:phase}.

Since the eh interlayer Coulomb potential is attractive, the free exciton model always leads to a negative binding energy no matter how weak the binding potential is or how dense the excitons are.
On the other hand, the eh plasma energy for small $r_s$ is positive since the kinetic energy dominates.
Thus, within our approximations, the exciton gas is the stable ground state at small $r_s$ independent of the value of $d$.  Although this seems to be an artifact of our approximation scheme, where exciton-exciton interactions are ignored, we believe that this result is correct since the attraction between the interlayer electrons and holes cannot be screened away giving the excitons always a negative energy (which could be very small) whereas the eh plasma must have a positive energy at small $r_s$ arising from the kinetic energy contribution. Of course the energy difference between the exciton gas and the eh plasma could be extremely small, leading to Saha ionization of the excitons into an eh plasma at finite temperatures.  But our current theory applies at $T=0$, where we believe that excitons are the stable ground state with a lower energy than the eh plasma even for very small $r_s$.

For general $r_s$ and d, however, our theory predicts a stable eh plasma around $r_s\sim2$ because of the minimum in the plasma energy around this $r_s$ value, but for very large $r_s$, again the exciton gas appears to be the preferred ground state of the system for arbitrary $d$.  We speculate that many body corrections ignored in our theory would most likely stabilize the exciton system for all $r_s$ and $d$ since we see no particular reason for the comparative stability to show nonmonotonicity as a function of $r_s$ and $d$.  But our theory for the plasma energy is accurate only for small $r_s$ whereas our theory for the exciton gas is accurate only for large $r_s$, and the intermediate $r_s$ regime, where we are finding a stable plasma phase is not a theoretically reliable regime for our approximations since both the plasma and the exciton energies are inaccurate at intermediate $r_s$ values in our approximation scheme.

\section{Conclusion}
By comparing the ground state energies of an electron-hole bilayer system assuming that it is an eh plasma and that it is an exciton gas, we have calculated the qualitative zero-temperature phase diagram under various assumptions of the screening of the Coulomb interaction.
We find that the exciton gas is favored at small $d$ if the unscreened interaction is used in the calculation of the exciton binding energy, while eh plasma is favored even at small $d$ if the statically screened Coulomb interaction is used.
For the dynamically screened excitons, we find that the result depends on which energy scale we use for the external frequency $\omega$.
In particular, the $\omega=-E_X$ and the $\omega=E_p$ case looks qualitatively similar to the unscreened situation, while the $\omega=E_F$ case appears to be similar to the statically screened case.
It is interesting that the energy scale of the dynamic screening is able to tune between unscreened (with stable excitons) and statically screened cases (without stable excitons). This suggests that nontrivial many-body interaction physics plays a key role in deciding the stability of the exciton gas, and that
it is not trivial to assume that the formation of exciton is always preferred in such bilayer system, and a more accurate model is required to settle down the problem of exciton formation.

We also find that within our approximation schemes, the exciton gas is quite generally stable for high (small $r_s$) and low (large $r_s$) densities, but the plasma phase may be stable for intermediate $r_s$ although we have argued that this may be an artifact of our theory being inaccurate for both the plasma and the exciton energetics in the intermediate $r_s$ regime.

We note that the exciton gas, being a gas of bosons, would form a superfluid excitonic condensate, producing a spontaneous interlayer coherence, which we do not consider at all in our theory where no symmetry breaking is allowed in the ground state.   The formation of such an interlayer electron-hole superfluid involves a condensate energy which further lowers the exciton energy, and it is possible that the inclusion of this condensation energy would automatically make the exciton system more stable at $T=0$ for all values of $r_s$ and $d$ (even if the exciton gas is higher in energy than the plasma for intermediate $r_s$ and $d$, as we find in the current approximations).  The superfluid formation is beyond the scope of the current work, where we only consider and compare the energetics of two independent limiting scenarios: an exciton gas and an electron-hole plasma.

\section*{Acknowledgment}

This work is supported by the Laboratory for Physical Sciences.
 
\bibliographystyle{apsrev4-2}
\bibliography{references}

\end{document}